\documentclass[%
reprint,
superscriptaddress,
 amsmath,amssymb,
 aps,
]{revtex4-1}

\usepackage{graphicx}
\usepackage{longtable}
\usepackage{enumerate}
\usepackage{hyperref}
\usepackage{multirow}

\usepackage{xcolor}

\begin{document}



\title{Complex networks approach to curriculum analysis and subject integration: a case study on Physics and Mathematics
}
\author{Paula Tuz\'on}
\email{paula.tuzon@uv.es}
\affiliation{Science Education Department, University of Valencia. Avda. Tarongers 4, 46022 Valencia, Spain}

\author{Antoni Salv\`a Salv\`a}
\affiliation{IES Binissalem, Binissalem, Spain}
\affiliation{Department of Chemistry, University of the Balearic Islands, Ctra. Valldemossa, Km 7.5, 07122 Palma, Spain}

\author{Juan Fern\'andez-Gracia}
\email{juanf@ifisc.uib-csic.es}
\affiliation{Instituto de Física Interdisciplinar y Sistemas Complejos IFISC (CSIC-UIB), Campus UIB, 07122, Palma de Mallorca, Spain}


\begin{abstract}
This paper presents a methodological approach based on the use of complex networks to analyze the structure and content of curricula. We analyze the concept network built from the final year of a particular high school Physics curriculum, as well as that of Mathematics. We examine the most central nodes in each case, the community structures (coherent units or groupings), and the changes that occurred when the network was considered in isolation or integrated with Mathematics. The results show that the integrated Physics and Mathematics network has a higher average degree compared to the individual networks, driven by numerous interdisciplinary connections. The modularity analysis indicates similarities with the original curriculum layout, but also interesting differences that may suggest alternative ways of organizing the content. The differences between separated and integrated networks also highlights the prominence of certain key concepts. 
\end{abstract}


\maketitle


\section{Introduction}

In recent years, educational approaches that move beyond traditional subject divisions have gained traction~\cite{Hacker1985,Lang2000,Kazeni2008}. These approaches aim to be more integrative, emphasizing the interaction between subjects. In the context of the sciences, this is evident in concepts such as STEM education~\cite{Zulkifli2022}, integrated science \cite{Harrell2010}, inquiry-based learning~\cite{Furtak2012}, and deep learning~\cite{Kovac2023}. Consequently, contemporary curricula increasingly allocate time for project-based learning and procedural skill development. This shift is closely aligned with the nature of science itself~\cite{Lederman2002,Lederman2019}, which, despite being highly specialized, often draws upon knowledge from other disciplines. Such interdisciplinary integration enables deeper exploration and can drive more significant advancements within specialized fields.

However, this integration must be approached with care. Exploring connections between subjects should not result in superficial understanding. Instead, the objective is to determine which elements from various subjects, when combined, lead to deeper and/or more effective learning. Achieving this is not always straightforward, as interdisciplinary integration can sometimes fail to meet pedagogical objectives~\cite{Capps2013,McComas2020,Takeuchi2020}. This may be partly due to the lack of a clear methodological proposal regarding which subjects or parts of subjects should be integrated and how.

In this context, network science can offer valuable insights. By representing topics as nodes and their epistemological relationships as links, network analysis can reveal which connections are more robust, which nodes are more central, and whether there are clusters of concepts that are best reinforced within a single subject or through interdisciplinary integration.

The application of network science to education is not a novel concept and has been explored in various studies \cite{Koponen2020}. These methods not only facilitate the visualization of complex relationships but also enable the quantification of indirect effects that traditional statistical methods may fail to capture. There are several studies on the coherence of content blocks through the analysis of community structures within students' concept networks and the centrality of their nodes. Various examples of the use of concept networks are the following: Some studies about general \cite{Sayama2017} or specific topics such as energy and electrostatics \cite{Podschuweit2020,Thurn2020} use these networks to evaluate and understand students' ideas and misconceptions. They are also used to evaluate learning processes \cite{Siew2020} or inferential reasoning chains of physics concepts \cite{Speirs2024}. In \cite{Kubsch2020}, they establish a correlation between the coherence of concept networks and the practical application of knowledge, while in \cite{Schwab2017} teaching sequences are personalized by considering the complex relationships between concepts. Related to concept networks, there are also studies analyzing students' answers to surveys. For example, in \cite{Riihiluoma2024} the authors analyze students' understanding of different approaches to quantum mechanics by examining the connections between responses to a questionnaire. Also multiple choice surveys about physics concepts provide interesting information about students' misconceptions related to other variables \cite{Stewart2021,Wells2020,Wheatley2022}. 
Apart from concept networks, network theory in education has been used to evaluate, for example, classmates networks and students' academic gain \cite{Zwolak2018,Traxler2018}, considering different contexts such as teaching mode (online vs. in-person) \cite{Pulgar2023} or teaching approach (active learning) \cite{Traxler2020}. Similarly, authorship networks in publications within the field have also been analyzed \cite{Anderson2017}, and social practices between instructors and students have been described using network theory \cite{Olsen2023}.

In this work, we present a methodological approach based on concept network analysis to evaluate curricular proposals. This methodology allows us to address several questions: Given the contents present in a specific curriculum and its epistemological relationships, which are the most important concepts in the network? Are the content blocks proposed by the curriculum the most coherent way to teach them? If we consider the content of two or more subjects together, do the same blocks or groupings make sense, or do other, more coherent groupings emerge? Are the key concepts equally important or central when a subject is taught independently compared to when it is integrated with another?

To showcase our methodological approach, we will apply the analysis to the contents of a Physics curriculum within a specific course and compare the resulting concept networks when Physics is studied independently versus alongside Mathematics. The epistemological connection between Physics and Mathematics is well established, and a comparative analysis of their respective concept networks—whether studied independently or in combination—yields valuable insights on how to combine subjects and how the combination of those change the importance of certain contents. In a context of competence-based learning, this knowledge is specially relevant for a proper curricular design approach.

\subsection{Basic concepts about complex networks}

To facilitate the reading of the article, we will briefly review some concepts in complex networks that are particularly relevant to this study~\cite{Newman2010}.

A network is a collection of nodes that are linked by edges. Mathematically it is defined as a set of nodes $\{v_i, i=1,N\}$, where $N$ is the number of nodes, and a set of edges $\{e_i, i=1,M\}$, where $M$ is the total number of edges. Each edge $e_i$ is formed by a pair of nodes, which it is connecting, {\it e.g.} $e_i = (v_j, v_k)$. Networks can be directed or undirected, depending if the edges have a prescribed direction or not. The {\it density} $\rho$ of a network is defined as the number of edges divided by the total number of edges that could be formed in the network, $\rho = 2M/N(N-1)$ (without the factor 2 for directed networks). A {\it path} on a network is a succession of non-repeated nodes for which there exists an edge between any two consecutive nodes. The length of the path equals the number of edges forming it. If a path can be constructed between any pair of nodes it is said that the network is {\it connected}. Otherwise the network consists of {\it components}, which are groups of nodes that form subnetworks that are connected. For a connected component one can define its {\it diameter} as the longest shortest path between any two nodes in it. The average length of all shortest paths is usually referred to as the {\it average path length}. Another typical measure in complex networks is the {\it clustering coefficient}, which, for a node, measures the proportion of edges between its neighbors that are present, from all the possible edges that could be formed. Typically this is averaged over all the nodes in the network to compute the average clustering coefficient.

Centrality measures describe the importance of nodes in a network in different ways. There are many centralities, but in this work we are going to concentrate on three: degree, closeness and betweenness centrality. The \emph{degree} is the number of links connected to a node and is the most basic centrality measure. It is a local measure that does not require knowledge of the global structure of the network. Typically, a highly connected node is considered very important or central to the network. Nodes with disproportionately large degrees in the network are referred to as hubs. In content networks, nodes with higher degrees represent content that can be related to many other contents, either due to their cross-disciplinary nature, their broad applicability, or because they combine results from different contents. \emph{Closeness} is the inverse of the average length of the shortest paths that connect a node to the rest of the nodes in the network. In this way, a node that is close to the rest of the nodes will be considered central. This measure requires global knowledge of the network. In content networks, this centrality measure will be able to identify the content that can be more easily related to the rest of contents on a global scale. \emph{Betweenness} is the number of shortest paths between all pairs of nodes in the network that pass through the node under study. This measure also requires global knowledge of the network. It can inform us about which content acts as a bridge between groups of content. In directed networks, we can measure analogous centralities to those in undirected networks. For degree centrality, we distinguish between \emph{in-degree} (the number of edges pointing toward a node) and \emph{out-degree} (the number of edges emerging from a node). Closeness centrality can be assessed in two ways: on the original network, where it measures how easily a node can reach all other nodes, or on the reversed network, where it indicates how easily a node can be reached from any other node. Betweenness centrality naturally extends to directed networks by considering only paths that follow the directionality of edges.

Networks can be described not only at a microscopic level (node-to-node) or at a macroscopic level with aggregate measures for the entire network, but they can also be studied at intermediate levels of organization. In fact, many networks reveal their richness at this level \cite{Boccaletti2006,Milo2002}. This is commonly referred to as the mesoscopic level and is studied through community structure. A community in a network is a set of nodes that are much more connected to each other than to the rest of the network. Different community detection algorithms can be defined \cite{Khan2017,Javed2018}, and it may be necessary to study which one will work best given the nature of the network under consideration \cite{Fortunato2016}.



\section{Methodology}

By viewing curricular contents as nodes and their dependencies as edges, network theory offers a powerful framework for organizing and sequencing educational content. This approach allows educators to better understand how different topics are interrelated.

\subsection{Official contents of Physics and Mathematics}

To illustrate our approach, we focus on the content of two subjects, Physics and Mathematics, as they are displayed in the official curriculum, i.e., the particular law where the basic knowledge or concepts to be taught are established. The relations between these curricular concepts are the subject of our data collection. We focus on a specific level, which is the final year of general upper secondary education, ``\emph{2º de bachillerato}'', according to the International Standard Classification of Education~\cite{isced}. In Spain, education is regulated at the national level by an organic law that governs public and private education across primary, secondary, post-secondary and vocational 
education~\cite{LOMLOE}. Based on this organic law, decrees outlining the minimum educational requirements are established, particularly for secondary education~\cite{Curriculum}. These decrees define the basic knowledges, i.e. contents, that each subject must cover. 
We focus on the contents of Physics and Mathematics. The current curriculum is less centered on content, as the present law emphasizes a competency-based approach~\cite{FernandezGracia_Competences}, as compared to the previous curriculum that was centered around contents. We rely on the content definitions from the previous curriculum decree~\cite{Curriculum_old}, as we created the network before the change in legislation. Nevertheless, both curricula cover the same key concepts with minimal variation in their structure.
 
\begin{figure*}[tbh]
\includegraphics[width=7cm]{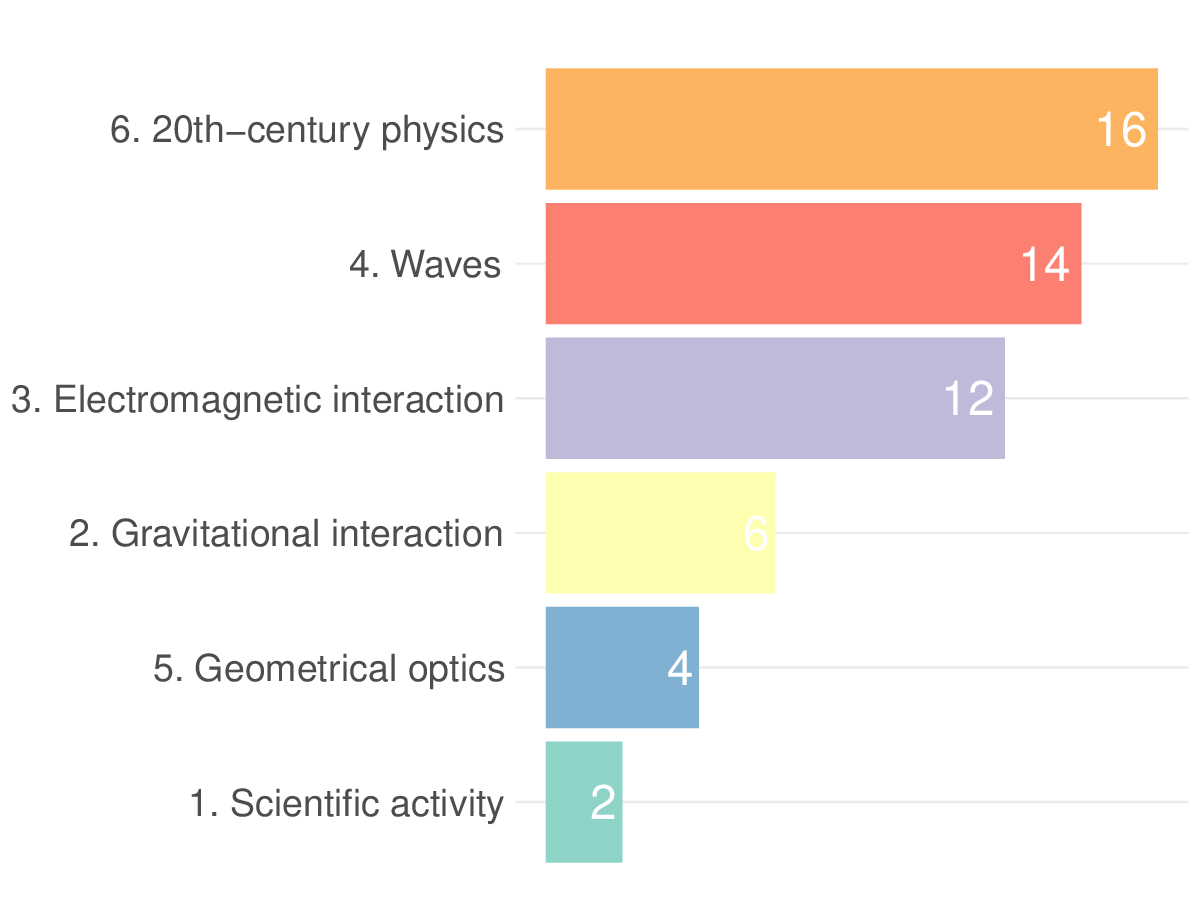}
\includegraphics[width=7cm]{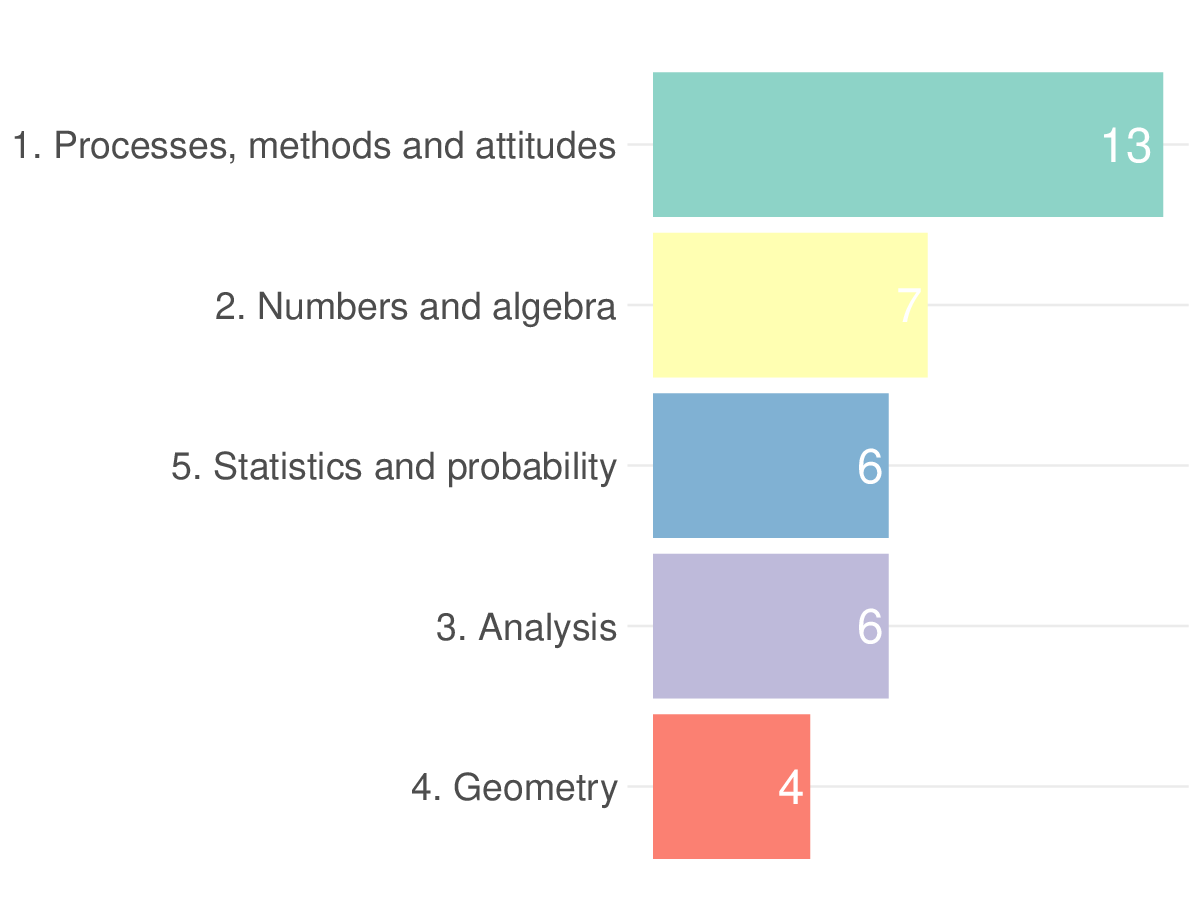}
\caption{\label{contents} Physics (left panel) and Mathematics (right panel) curricular blocks of contents and amount of contents per block.}
\end{figure*}

The subject of Physics, according to this curriculum, consists of 54 contents distributed in 6 blocks or units~\cite{Curriculum_old}. Mathematics consists of 36 contents divided into 5 blocks. The titles of the blocks can be read in the Figure \ref{contents}, which also shows the number of contents for each block.
\subsection{Contents network construction}

The relationships between the contents were manually reviewed by the authors of this article. Two contents were connected whenever they overlapped in concepts or methodologies. There are three types of relationships that may occur between two contents: 1) a connection with dependency, meaning one content is required to understand the other; 2) a connection without dependency, meaning both contents are related but the order in which they are taught does not affect comprehension; and 3) a disconnection, meaning there is no significant relationship between the contents in terms of the learning process. These relationship types allow for the construction of concept networks, which can help propose content blocks, forming the basis for learning units, learning situations or subjects.

For $N$ contents, the total number of relationships to check is $N(N - 1)/2$. In our case, with 90 contents in total, this results in 4005 relationships to verify. To carry out this task systematically, we developed an automatic data collection method. This method presents two contents and prompts us to assess their dependency on one another. The collected data, along with a log of previously checked relationships, is saved, allowing us to continue verifying relationships across multiple sessions without the risk of overlooking any. To minimize potential bias, the data collection method does not indicate the specific content block to which the contents belong.

We constructed both directed and undirected networks and used one or the other depending on the analysis. In the undirected network, we treated both types 1) and 2) (connections with and without dependency) as equivalent, replacing directed relationships (type 1) with undirected ones and disregarding the order in which contents should be taught. In the directed network, we maintained the original directionality and converted undirected relationships to bidirectional directed edges.

\subsection{Identifying central concepts}

We investigated three different types of centrality on the undirected networks. A local measure, which is the degree of each content, i.e. the number of contents related to the same; and two global ones, namely the closeness centrality, which measures how close the rest of the nodes are to one another, and the betweenness centrality, which measures for each content the percentage of all the shortest paths between nodes in the network that pass through one. On the directed version of the network we used 5 different variations of the same centralities, namely in-degree, out-degree, closeness on the original and reversed network and betweenness.

These centrality measures allow us to create rankings of concepts in terms of importance. The rankings are compared to detect whether there are different central concepts depending on the definition of centrality. The rankings are further compared when the subjects are considered in isolation vs. when they are considered together.

\subsection{Community structure}

The analysis of communities has been done on the undirected networks with own code written in Python, using the existing implementation of the Infomap algorithm \cite{Rosvall2009,Edler2020}. This algorithm uses random walkers on the network to infer its community structure. A random walker is an agent that can move from one node to another by randomly following the links in the network. If the network has a pronounced community structure, the random walker will remain trapped within each community for a long time before jumping to another. The groupings that arise from the community detection analysis are compared with the grouping into blocks when the subjects are isolated. 

\subsection{Network visualization}

The Cytoscape program was used to visualize the networks \cite{cytoscape}. A mixture of own code written in Python and Cytoscape functionalities was used for the analysis. 

\section{Results}

\subsection{Physics subject}

\begin{figure*}[tbh]
\includegraphics[width=0.7\textwidth]{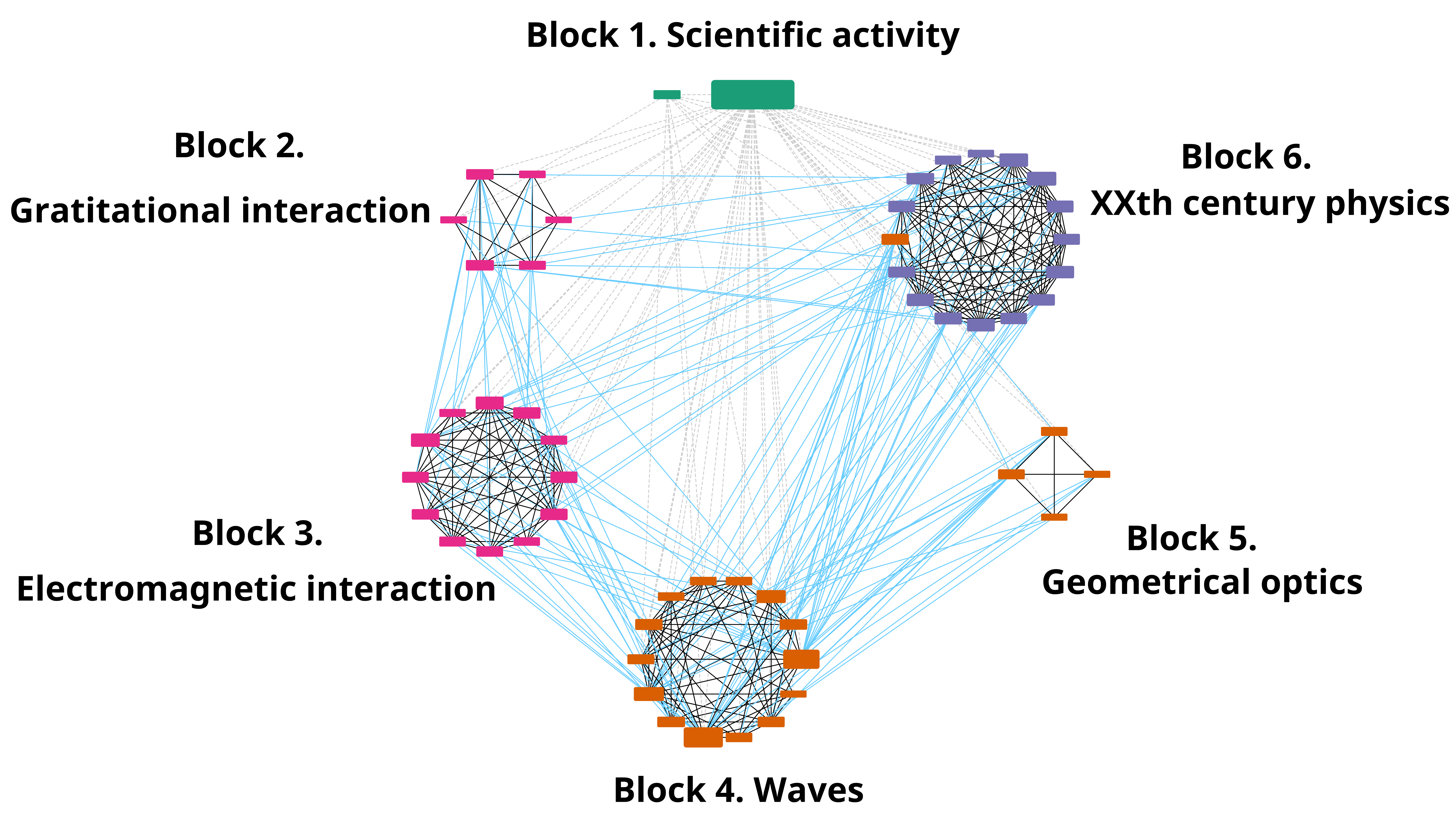}
\caption{\label{physnet} Physics concept network. The contents are grouped into blocks. The links to block 1 are shown in gray and with a dotted line to aid in visualizing the rest of the network. The remaining links are represented by solid lines, with black lines connecting contents within the same block and blue lines connecting contents from different blocks. The height of each content's symbol is proportional to its degree, while the width corresponds to its betweenness centrality. The color of the nodes represents the module they belong to, according to the community analysis.}
\end{figure*}

The network of the Physics subject concepts is represented in Figure~\ref{physnet}, where contents are grouped into blocks. For this subject we found 409 connections between contents out of the 1431 possible. This means a density of 0.29 and that, on average, each piece of content is connected to 15.15 other pieces of content. The network is composed of only one connected component, which means that a path can be found on the network to go between any two contents. These characteristics denote the non-linearity and complexity of the relationships between the contents. In fact, the diameter of the network is 2, meaning that two contents that are not connected, are always connected through a third party. As we will see later, this third is typically a content of the first block, very central as it is more transversal. The average path length is 1.71 and the clustering coefficient is 0.65, a fairly high value. This means that in 65$\%$ of cases two contents that are connected to a third party are also connected to each other. 

\begin{figure*}[tbh]
\includegraphics[width=\textwidth]{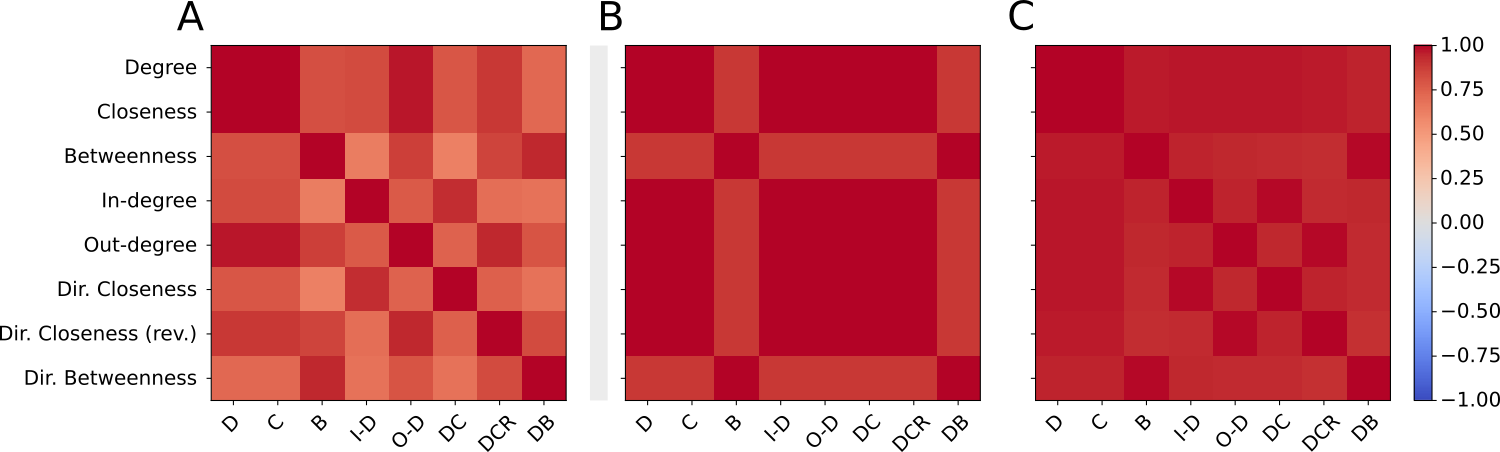}
\caption{\label{allcorr}
Spearman rank correlation of the pairs of centrality measures on the undirected and directed networks for Physics ({\bf A}), Mathematics ({\bf B}) and the joint network ({\bf C}). The first three centralities are measured on the undirected network: degree (D), closeness (C) and betweenness (B). The other five measures are calculated on the directed version of the networks: in-degree (I-D), out-degree (O-D), closeness on the original and on the reversed directed network (DC and DCR), and betweenness (DB).
}
\end{figure*}

The eight centrality measures we calculated—degree, in-degree, out-degree, closeness (on the original and reversed networks), and betweenness (for both directed and undirected cases)—are highly correlated, with a minimum Spearman correlation of 0.61 ($p<0.001$, Figure \ref{allcorr}). This strong correlation suggests that the rankings derived from these measures are largely consistent, allowing us to rely on degree centrality, the most parsimonious measure, to assess content importance without requiring full knowledge of the network structure.

Directionality does influence the rankings, but not to a great extent. This is confirmed by the fact that, regardless of the specific centrality measure used, at least five concepts are shared between any pair of measures in their respective top ten rankings.

The ten most central contents according to degree can be found in Table \ref{phys_central}. The most central content is  ``Own strategies for scientific activity'' in block 1, ``Scientific activity''. The next two contents are in block 4 ``Waves'' and are related to electromagnetic waves. Among the other 10 central contents, there are three of block 6 ``Physics of the XX century'', two of block 4, ``Waves'', and two of block 3, ``Electromagnetic interaction''. Centralities of all the contents can be checked visually by means of Figure~\ref{physnet}. \

\begin{table*}
\footnotesize
\caption{The ten most central contents of the Physics subject.\label{phys_central}}
\begin{ruledtabular}
{\setlength{\tabcolsep}{0.5em}
\begin{tabular}{p{4cm}p{6.5cm}p{0.5cm}p{1cm}p{1.6cm}}
\textbf{Block} & \textbf{Content} & \textbf{Degree} & \textbf{Closeness} & \textbf{Betweeness} \\
\hline 
1. Scientific activity & Specific strategies to scientific activity & 53 & 1.0 & 0.338 \\
4. Waves & Electromagnetic waves & 34 & 0.736 & 0.077 \\
4. Waves & Nature and properties of electromagnetic waves & 32 & 0.716 & 0.06 \\
6. 20th-century physics & The four fundamental interactions of nature: gravitational, electromagnetic, strong nuclear and weak nuclear & 21 & 0.624 & 0.019 \\
4. Waves & Wave phenomena: interference and diffraction, reflection and refraction & 21 & 0.624 & 0.023 \\
3. Electromagnetic interaction & Electric field & 20 & 0.616 & 0.015 \\
6. 20th-century physics & Quantum physics & 20 & 0.616 & 0.014 \\
4. Waves & The electromagnetic spectrum & 19 & 0.609 & 0.018 \\
3. Electromagnetic interaction & Magnetic field & 19 & 0.609 & 0.01 \\
6. 20th-century physics & Limitations of classical physics & 19 & 0.609 & 0.008 \\
\end{tabular}}
\end{ruledtabular}
\end{table*}

To analyze the community structure we have discarded from the network the contents of the 1st block, so we consider that they are transversal and will deal with all the contents. The Infomap algorithm identifies three content modules. The first module includes the 2nd and 3rd blocks, ``Gravitational interaction'' and ``Electromagnetic interaction''. The second module includes the blocks ``Waves'' and ``Geometric optics'', as well as a content of the 6th block, ``Applications of quantum physics. The laser''. Finally, the 3rd module contain the rest of the contents of the 6th block.

\subsection{Mathematics subject}

Since we will analyze the Mathematics subject to evaluate its network of concepts alongside that of Physics, we will also include the analysis of the Mathematics network of concepts in isolation for completeness.

\begin{figure*}[tbh]
\includegraphics[width=0.7\textwidth]{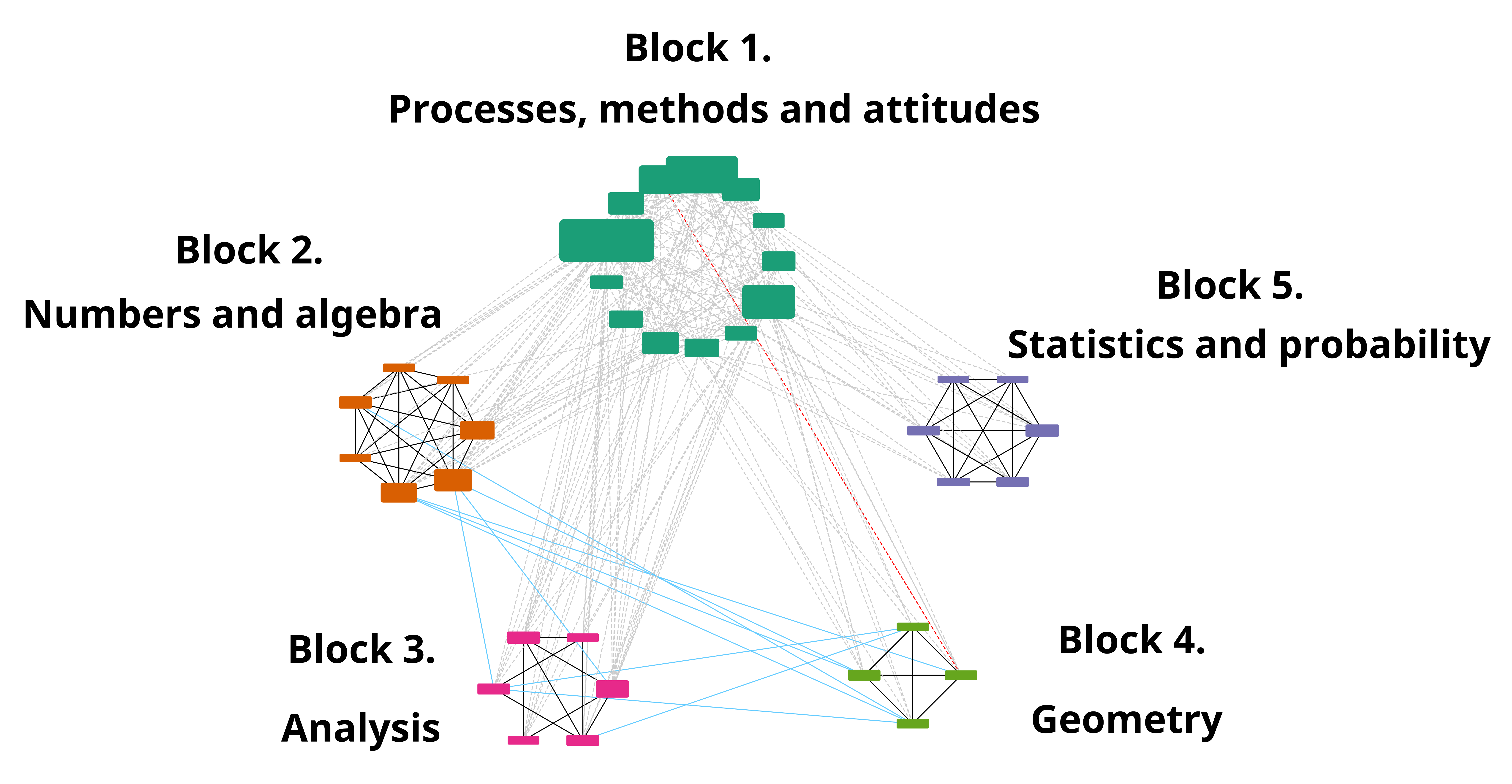}
\caption{\label{mathsnet} Mathematics concept network.}
\end{figure*}

In the case of Mathematics, the concept network is represented in Figure~\ref{mathsnet}. We have found 249 connections out of 630. This means a density of 0.395 and that, on average, each piece of content is connected to 13.83 other contents. The network is composed of only a single component. We see again that the contents form a complex non-linear network, with many connections between different contents of different blocks. The diameter of the network is of length 2 and the length of the average path is 1.6. As in the Physics subject, we see that between two contents there is always a connection or a connection through a very central 3rd content. The clustering coefficient is 0.74, a very high value. This means that in 74$\%$ of cases two contents that are connected to a third party are also connected to each other.

The eight centrality measures show a strong and consistent correlation, with the exception of betweenness (both in the directed and undirected cases), which differs slightly from the others (Figure \ref{allcorr}). This is reflected in its lower Spearman rank correlation of 0.89 ($p<0.001$) with the remaining measures. Despite this, the overall agreement among centrality rankings remains high, reinforcing the reliability of degree as a representative measure. The consistent relationships between these measures suggest that the network structure inherently supports a shared notion of centrality across different definitions.

Again we will use the degree to rank the nodes, as the most simple measure of centrality. Only betweenness differs in the top ten, but sharing 9 contents. The 10 most central contents of the network are dominated by contents of the 1st block, ``Processes, methods and attitudes in mathematics'', which are very transversal. We also find two contents of the 2nd block, ``Numbers and algebra'', related to matrix applications. We can consult these 10 contents in Table \ref{math_central}.

\begin{table*}
\footnotesize
\caption{The ten most central contents of the Mathematics subject.\label{math_central}}
\begin{ruledtabular}
{\setlength{\tabcolsep}{0.5em}
\begin{tabular}{p{4cm}p{6.5cm}p{0.5cm}p{1cm}p{1.6cm}}
\textbf{Block} & \textbf{Content} & \textbf{Degree} & \textbf{Closeness} & \textbf{Betweeness} \\
\hline 
1. Processes, methods and attitudes & Confidence in one's own abilities to develop appropriate attitudes and face the difficulties inherent in scientific work & 35 & 1.0 & 0.209 \\
1. Processes, methods and attitudes & Use of technological tools in the learning process to: a) Collect data [...] b) Elaborate and create graphic representations [...] c) Understand geometric or functional properties [...] d) Design simulations and make predictions [...] e) Prepare reports [...] f) Communicate and share mathematical information [...] & 331 & 0.897 & 0.135 \\
1. Processes, methods and attitudes & Graphical language, algebraic, other forms of representation of arguments & 28 & 0.833 & 0.071 \\
1. Processes, methods and attitudes & Solutions and/or results obtained: consistency of the solutions with the situation, systematic review of the process, other forms of resolution, similar problems, interesting generalizations and particularizations & 24 & 0.761 & 0.04 \\
1. Processes, methods and attitudes & Practice of mathematization and modeling processes, in contexts of reality and in mathematical contexts & 20 & 0.7 & 0.019 \\
2. Numbers and algebra & Application of matrix operations and their properties in solving problems taken from real contexts & 19 & 0.686 & 0.022 \\
1. Processes, methods and attitudes & Strategies and procedures put into practice: relation to other known problems, modification of variables, assuming the problem is solved & 19 & 0.686 & 0.018 \\
1. Processes, methods and attitudes & Carrying out mathematical research based on contexts of reality or contexts of the world of mathematics & 19 & 0.686 & 0.015 \\
2. Numbers and algebra & Matrix representation of a system: discussion and solution of systems of linear equations. Gauss method. Cramer's rule. Application to problem solving & 17 & 0.66 & 0.016 \\
1. Processes, methods and attitudes & Planning the problem solving process & 17 & 0.66 & 0.007 \\
\end{tabular}}
\end{ruledtabular}
\end{table*}

Regarding the structure of communities, we find that they completely match with the blocks proposed in the curriculum. This analysis has been done excluding the 1st block of the subject, for the same reasons that has been done with the subject of Physics: since they are transversal contents, having them in the network can provide false information
of connectivity between contents in the Infomap community search algorithm.

\subsection{Physics and Mathematics integration}

\begin{figure*}[tbh]
\includegraphics[width=0.7\textwidth]{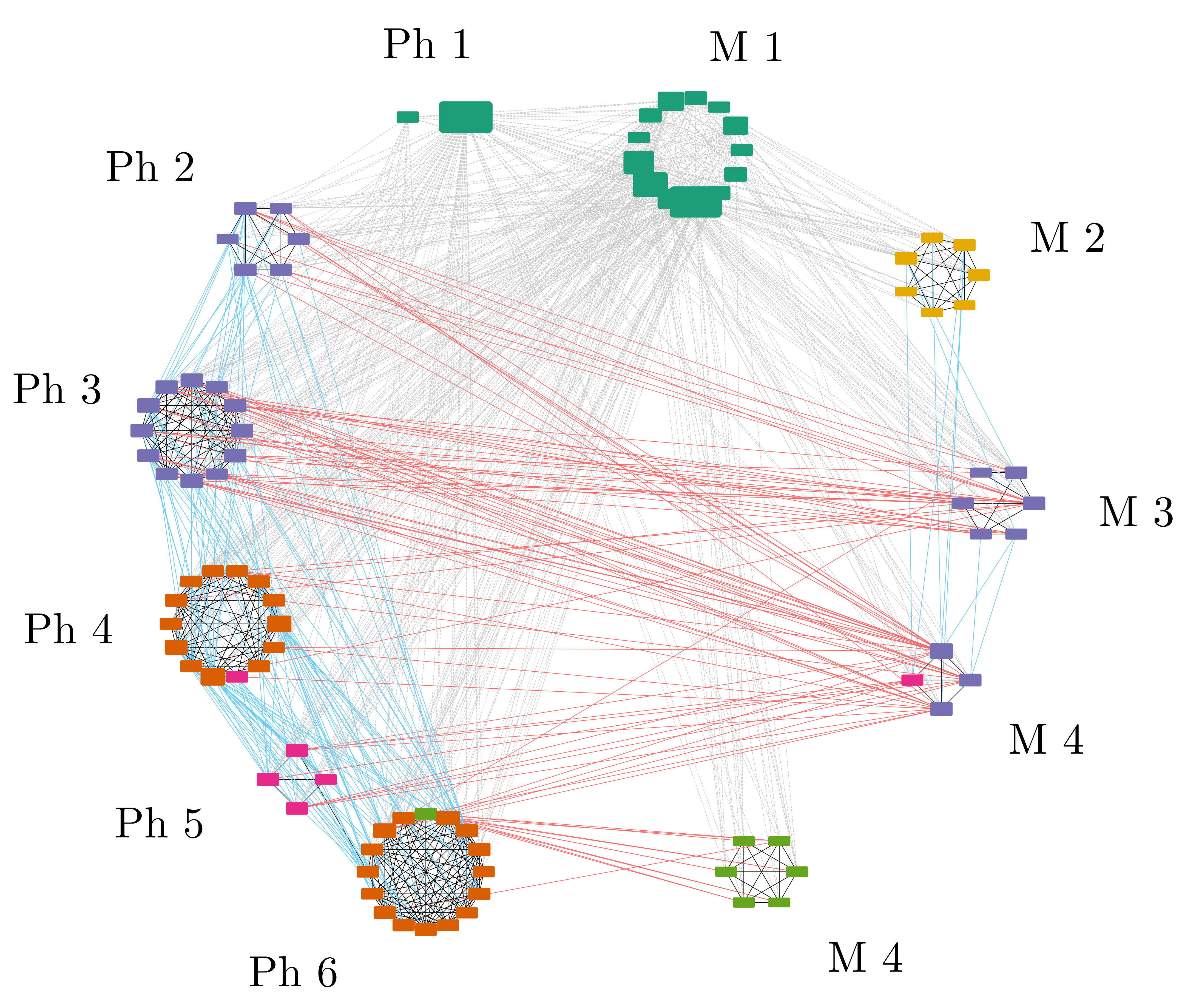}
\caption{\label{physmathsnet} Physics and Mathematics concept network. On the left side of the figure, we find the contents of physics, while on the right side are those of mathematics. The contents are grouped into blocks. Each block is identified as ``Ph'' or ``M'', depending on whether it is Physics or Mathematics, followed by the corresponding number. The links to block 1 of each subject are shown in gray and with a dotted line to aid in visualizing the rest of the network. The remaining links are represented by solid lines: black lines connect contents within the same block, blue lines connect contents from different blocks within the same subject, and red lines connect contents from different subjects. The height of each content's symbol is proportional to its degree, while the width corresponds to its betweenness centrality. The color of the nodes represents the module they belong to, based on the community analysis.}
\end{figure*}

The complete network of Physics and Mathematics contents comprises 90 items, with 1016 connections identified out of a possible 4005 (see Figure \ref{physmathsnet}). This results in an average degree of 22.58, meaning each content is connected to 22 other contents on average. The network density is 0.25. The connections are structured such that the network is a single component. Compared to when the subjects are treated as separate networks, the average number of connections increases due to the numerous links between contents of both subjects; however, the density is lower than that of either subject’s individual network. This indicates a complex and non-linear structure. The network diameter is 2, and the average path length between pairs of contents is 1.75. Additionally, the phenomenon induced by the central methodological contents from the first blocks persists, ensuring that all contents can be connected with at most one intermediary content, further highlighting the network’s tight interconnectedness.

There is a significant amount of links between the two subjects (red links in Figure \ref{physmathsnet}). It is quite significant that the 5th block of Mathematics, ``Statistics and probability'', has more connections with Physics contents than Mathematics. More specifically, it is related to quantum physics and its probabilistic interpretation.\\

As observed in the individual content networks, the eight centrality measures in the global network of both subjects remain highly correlated (Figure \ref{allcorr}). The results reflect a mixture of those obtained for the separate networks, with some variation in the correlations between centrality measures. However, these variations are less pronounced than in the Physics network, with a minimum Spearman rank correlation of 0.91 ($p<0.001$). This consistency further supports the robustness of the centrality rankings across different network structures.

Table \ref{physmath_central} shows the 10 most central contents of the complete network, ordered according to the degree. When using other centrality measures the top ten nodes remain the same except for betweenness in the undirected case. Still, this ranking retains 9 of the top ten nodes. The first thing we see is that the list is dominated by transversal contents of the 1st block of the two subjects. We also note that the two contents about electromagnetic waves of the Physics subject continue to be one of the most central. There are two contents that have not appeared before, both from the Mathematics subject: ``Deductive and inductive reasoning'' from Block 1 in the 6th place and ``Vectors in three-dimensional space. Scalar, vector and mixed product. Geometrical meaning'' from Block 4 in 10th place.

\begin{table*}
\footnotesize
\caption{The ten most central contents considering together Physics and Mathematics content network.\label{physmath_central}}
\begin{ruledtabular}
{\setlength{\tabcolsep}{0.5em}
\begin{tabular}{p{4cm}p{6.5cm}p{0.5cm}p{1cm}p{1.6cm}}
\textbf{Block} & \textbf{Content} & \textbf{Degree} & \textbf{Closeness} & \textbf{Betweeness} \\
\hline 
Ph1. Scientific activity & Specific strategies to scientific activity & 89 & 1.0 & 0.179 \\
M1. Processes, methods and attitudes & Confidence in one's own abilities to develop appropriate attitudes and face the difficulties inherent in scientific work & 88 & 0.989 & 0.17 \\
M1. Processes, methods and attitudes & Use of technological tools in the learning process to: a) Collect data [...] b) Elaborate and create graphic representations [...] c) Understand geometric or functional properties [...] d) Design simulations and make predictions [...] e) Prepare reports [...] f) Communicate and share mathematical information [...] & 66  &  0.795 & 0.073   \\
M1. Processes, methods and attitudes & Graphical language, algebraic, other forms of representation of arguments & 61 & 0.761 & 0.049  \\
M1. Processes, methods and attitudes & Solutions and/or results obtained: consistency of the solutions with the situation, systematic review of the process, other forms of resolution, similar problems, interesting generalizations and particularizations & 48  & 0.685 & 0.028 \\
M1. Processes, methods and attitudes & Deductive and inductive reasoning & 44 & 0.664 & 0.027 \\
M1. Processes, methods and attitudes & Practice of mathematization and modeling processes, in contexts of reality and in mathematical contexts & 42 & 0.654 & 0.019  \\
Ph4. Waves & Electromagnetic waves & 39  & 0.64 & 0.017  \\
Ph4. Waves & Nature and properties of electromagnetic waves & 35  & 0.622  & 0.014 \\
M4. Geometry & Vectors in three-dimensional space. Scalar, vector and mixed product. Geometric meaning. & 31 & 0.605  & 0.009  \\
\end{tabular}}
\end{ruledtabular}
\end{table*}

The communities structure has been done again by excluding those contents of the 1st blocks of each subject. The algorithm finds 5 modules. These highly respect the block structure of the contents of the curriculum, grouping some of them. Mathematics Block 2,  ``Numbers and algebra'', remains as a module by itself. Blocks 2 and 3 of Physics (``Gravitational interaction'' and ``Electromagnetic interaction'') along with Blocks 3 and 4 of Mathematics (``Analysis'' and ``Geometry'') form another module, excluding the content ``Equations of the line and the plane in space'' of Block 4 of Mathematics. This content forms another module together with the contents of Block 5 of Physics,  ``Geometric optics'' and the content ``Dispersion. The color'' of Block 4 of Physics, ``Waves''. The rest of the contents of Block 4 of Physics,  ``Waves'', form a module with the contents of Block 6, ``Physics of the 20th century'', except for the content ``Probabilistic interpretation of quantum physics''. This content forms another module with the contents of Block 5 of Mathematics, ``Statistics and probability''.

\section{Discussion}

Our analysis demonstrates the utility of network theory in organizing and interpreting the relationships within and across educational content. The results show that the eight centrality measures yielded highly similar results, indicating that the network topology does not require separate centrality criteria for content ordering. This uniformity allows us to rely on degree as a reliable indicator of content importance, regardless also of directionality.

\subsection{Centrality Insights}

The primary purpose of Physics in the final year of high school is to connect the limitations of classical models with the basic aspects (at least qualitatively) of modern physics. This is consistent with the distribution of the most central concepts shown in the Table~\ref{phys_central}. In addition to the procedural aspect (``Specific strategies to scientific activity''), these are essentially: 1) waves, 2) fields, and 3) issues related to the introduction of modern physics. Concerning point 3, we find: fundamental interactions (beyond gravity and electromagnetism, i.e., the strong and weak interactions), quantum physics, and a topic called  ``limitations of classical models,'' which leads to the new insights of 20th-century physics. Waves, the concept of fields, and the general aspects of modern physics serve as the core around which the other concepts revolve in a pre-university physics course.

In the case of Mathematics, the most central concepts are mostly procedural (from the first block). The same happens in the joint network, where the most central concepts are dominated by the procedural contents of Mathematics, as we will discuss below. And those of Physics that either relate to the procedural part or deal with waves.

The ten most central contents in the networks were predominantly from the initial blocks of both subjects. These blocks contain highly transversal content, such as \emph{“Strategies specific to scientific activity”} or \emph{“Confidence in one’s own abilities to address challenges in scientific work”}. These contents are naturally central due to their generic and foundational nature, supporting a wide range of topics across the curriculum.

A second type of central content emerged from the analysis: contents that connect diverse concepts. For example, in Physics, Block 4’s \emph{“Electromagnetic Waves”} contents link to topics within their block, as well as to Block 3 (\emph{“Electromagnetic Interaction”}), Block 5 (\emph{“Geometric Optics”}), and Block 6 (\emph{“Physics of the 20th Century”}). These links highlight their broad interdisciplinary role, connecting concepts such as the photoelectric effect, radiation, and the relativistic Doppler effect. Similarly, \emph{“Limitations of Classical Physics”} from Block 6 illustrates the challenges of early 20th-century physics, connecting to a variety of content.

In Mathematics, central contents included operations with matrices, such as \emph{“Matrix representation of a system”} and \emph{“Gaussian method”}, which play a critical role in diverse mathematical and applied contexts. When Physics and Mathematics are integrated into a single network, additional connections emerge. For example, \emph{“Vectors in three-dimensional space”} from Mathematics becomes central due to its applications in Physics Blocks 2 and 3 (\emph{“Gravitational Interaction”} and \emph{“Electromagnetic Interaction”}), as well as in understanding waves and geometric optics. Another key mathematical content, \emph{“Deductive and inductive reasoning”}, also gains prominence when networks are integrated, reflecting its foundational role in building physical concepts like the relation between energy and orbital movement (content Ph2.5), or the Doppler effect (Ph4.6).

\subsection{Towards an Integrated Science Education}

From the integration of the Physics and Mathematics curricula, we have identified several relevant issues:

First, Physics Blocks 2 and 3, \emph{“Gravitational Interaction”} and \emph{“Electromagnetic Interaction”}, integrate well with Mathematics Blocks 3 and 4, \emph{“Analysis”} and \emph{“Geometry”}. This integration appears natural, as many of the mathematical tools required for understanding the physics blocks are explained in the mathematics blocks. Merging these blocks into mixed teaching units and learning situations could provide students with a more practical view of mathematics and, likely, a better understanding of physics.

Second, the importance of a given content, measured by its centrality in the network, can change depending on whether we consider the isolated network of a single subject or the integrated network of multiple subjects. For instance, when analyzing the joint network of Physics and Mathematics, \emph{“Vectors in three-dimensional space”} from Mathematics Block 4 became one of the ten most central topics. Adding Physics content revealed its significance, particularly for Physics Blocks 2 and 3. Similarly, the procedural skill \emph{“Deductive and inductive reasoning”} from Mathematics Block 1 gained prominence in the integrated network, highlighting its importance for building physical concepts like .

Finally, Mathematics Block 5, \emph{“Statistics and Probability”}, integrates with quantum physics content from Physics, providing greater context when the subjects are considered together. This integration emphasizes how interdisciplinary connections can help students develop a more complete and interconnected view of science.

These findings suggest that without collaboration and communication between teachers of different subjects, the true significance of certain topics might be overlooked. Indeed, cooperation and teamwork of teachers, as defined by the concept of Collective Teacher Efficacy, has been recognized as significantly positive influence related to student achievement~\cite{eells,Goddard2000}. By applying the approach presented in this paper to all science subjects at the secondary and other educational levels, we could propose a comprehensive framework for integrated science education. This would help identify the most relevant synergies and guide curriculum design to foster interdisciplinary learning.

\subsection{Implications for Teaching}

These findings suggest that centrality measures are effective tools for identifying two types of content: transversal contents that underpin the curriculum and conceptually rich contents that serve as hubs connecting multiple topics. Both types play crucial roles in the teaching-learning process. Transversal contents provide foundational skills, while conceptually rich contents reinforce multiple topics simultaneously, offering a holistic view of the subject. These contents should be prioritized and handled with special attention in curriculum design.

\subsection{Community Structure}

The community structure analysis revealed a modular organization that closely aligns with the content blocks in both subjects. In Physics, the Infomap algorithm grouped Blocks 2 and 3 (\emph{“Gravitational Interaction”} and \emph{“Electromagnetic Interaction”}), reflecting their shared tools and concepts (fields, forces, potentials). Blocks 4 and 5 (\emph{“Waves”} and \emph{“Geometric Optics”}) were grouped together with \emph{“Applications of Quantum Physics: Lasers”}. This alignment underscores the coherence of these blocks, as geometric optics involves electromagnetic waves, and lasers provide an ideal experimental tool. The remaining contents of Block 6 (\emph{“Physics of the 20th Century”}) formed a separate module, further reinforcing the thematic distinctiveness of modern physics.

In Mathematics, the community structure matched the block organization exactly, reflecting a well-defined internal consistency. Importantly, during data collection, the authors did not know the block assignments of the contents, highlighting the objectivity of the method in uncovering natural groupings.

\subsection{Integration of Physics and Mathematics}

The integrated Physics and Mathematics network demonstrated a higher average degree compared to the individual networks, driven by numerous interdisciplinary connections. However, the network’s density decreased, reflecting the broader scope of integration. Notable interdisciplinary connections include the dominance of Physics contents like \emph{“Electromagnetic Waves”} and the emergence of Mathematics contents such as \emph{“Vectors in three-dimensional space”}. These connections highlight opportunities for integrated STEM education that bridge disciplinary boundaries.

\subsection{Conclusion}

This study demonstrates that network analysis provides a robust framework for understanding curricular content. By identifying central and modular contents, this approach supports curriculum designers in prioritizing key topics and exploring interdisciplinary connections. While this work focuses on Physics and Mathematics, the methodology can be extended to other disciplines and educational levels to inform curriculum reform, guide curricular design approaches, and promote integrated learning experiences by means of teachers' cooperation.



\section{Acknowledgements}

JFG acknowledges funding from CSIC through the 2024ICT014 JUAN FERNÁNDEZ GRACIA project (Ayudas extraordinarias a la incorporación de científicos titulares OEP 2020-2021) and from project CEX2021-001164-M financed by MICIU/AEI /10.13039/501100011033.

\section{Data and code availability}

All data and code are publicly accessible through the GitLab repository in~\cite{repo_content_networks}

%

\end{document}